\documentclass[aps,prd,preprint,groupedaddress,showpacs]{revtex4}

\usepackage{epsfig}
\usepackage{graphics}

\usepackage{slashed}
\newcommand{\notq}{{\slashed{q}}}

\begin{document}

\preprint{DESY~08--194\hspace{12.4cm} ISSN 0418-9833}
\preprint{December 2008\hspace{14.95cm}}

\title{Open charm production at high energies and the quark Reggeization
hypothesis}

\author{\firstname{B.A.} \surname{Kniehl}}
\email{kniehl@desy.de}
\author{\firstname{A.V.} \surname{Shipilova}}
\email{alexshipilova@ssu.samara.ru}
\affiliation{{II.} Institut f\"ur Theoretische Physik, Universit\" at Hamburg,
Luruper Chaussee 149, 22761 Hamburg, Germany}

\author{\firstname{V.A.} \surname{Saleev}}
\email{saleev@ssu.samara.ru} \affiliation{Samara State University,
Academic Pavlov Street~1, 443011 Samara, Russia}

\begin{abstract}
We study open charm production at high energies in the framework of the
quasi-multi-Regge-kinematics approach applying the quark-Reggeization
hypothesis implemented with Reggeon-Reggeon-particle and
Reggeon-particle-particle effective vertices.
Adopting the Kimber-Martin-Ryskin unintegrated quark and gluon distribution
functions of the proton and photon, we thus nicely describe the proton
structure function $F_{2,c}$ measured at DESY HERA as well as the
transverse-momentum distributions of $D$ mesons created by photoproduction
at HERA and by hadroproduction at the Fermilab Tevatron.
\end{abstract}

\pacs{12.39.St, 12.40.Nn, 13.60.Hb, 13.85.Ni}

\maketitle

\section{Introduction}

The study of open charm production in high-energy lepton-hadron and
hadron-hadron collisions is considered as a test of the general applicability
of perturbative quantum chromodynamics (QCD) and also provides information on
the parton distribution functions (PDFs) of protons and photons.
The present analysis is to explore our potential to access a new dynamical
regime, namely the high-energy Regge limit, which is characterized by the
condition $\sqrt S \gg \mu  \gg \Lambda_{\rm QCD}$, where $\sqrt S$ is the
total collision energy in the center-of-mass (CM) reference frame,
$\Lambda_{\rm QCD}$ is the asymptotic scale parameter of QCD, and $\mu$ is the
typical energy scale of the hard interaction.
In the processes of heavy-quark production, one has $\mu\ge m$, where $m$ is
the heavy-quark mass.
In this high-energy regime, the contribution from partonic subprocesses
involving $t$-channel parton (quark or gluon) exchanges to the production
cross section can become dominant.
Thus, the transverse momenta of the incoming partons and their off-shell
properties can no longer be neglected, and we deal with {\it Reggeized}
$t$-channel partons.

The quasi-multi-Regge-kinematics (QMRK) approach \cite{QMRK1,QMRK2} is
particularly appropriate for this kind of high-energy phenomenology.
It is based on an effective quantum field theory implemented with the
non-Abelian gauge-invariant action, as suggested a few years ago
\cite{KTLipatov}.
Our previous analyses of charmonium and bottomonium production at the Fermilab
Tevatron \cite{PRD2003}
demonstrated the advantages of the high-energy factorization scheme over
the collinear parton model as far as the description of experimental data is
concerned.
These observations were substantiated for $B$-meson production at the Tevatron
in Ref.~\cite{TeryaevBB}, where the experimental data were again well
described using the Fadin-Lipatov effective Reggeon-Reggeon-gluon vertex
\cite{QMRK1}.
In Ref.~\cite{SVA08}, where the effective photon-Reggeon-quark vertex was
obtained and for the first time, the hypothesis of quark Reggeization was
successfully used to describe experimental data on single prompt-photon
production and on the proton structure functions $F_2$ and $F_L$.

The CDF Collaboration measured the differential cross sections $d\sigma/dp_T$
for the inclusive production of $D^0$, $D^\pm$, $D^{\star\pm}$, and $D_s^\pm$
mesons in $p\overline{p}$ collisions in run~II at the Tevatron as functions of
transverse momentum ($p_T=|\vec{p}_T|$) in the central rapidity ($y$) region
\cite{DTevn}.
These measurements were compared with theoretical predictions obtained at
next-to-leading order (NLO) in the collinear parton model of QCD
\cite{Nason,KKSS} taking into account quark and hadron mass effects, and it
was found that the latter improve the description of the data.

The differential cross sections $d\sigma/dp_T$ and $d\sigma/d y$ for
inclusive $D^{*\pm}$ and $D_s^{\pm}$ photoproduction measured by the
H1 \cite{H1HERA} and ZEUS \cite{HERADst,HERADs} collaborations at the
DESY HERA Collider were compared with NLO predictions in the collinear parton
model.
For $D^{*\pm}$ mesons, this was done in three approaches: the zero-mass
variable-flavor-number scheme (ZM-VFNS) \cite{mc01,mc02}, the
fixed-flavor-number scheme (FFNS) \cite{mc}, and the general-mass
variable-flavor-number scheme (GM-VFNS) \cite{Kramer:2001gd}.
The experimental results were found to generally lie above the NLO
expectations.
For $D_s^{\pm}$ mesons, the calculations were performed in the FFNS \cite{mc}
and in the model suggested  by Berezhnoy, Kiselev, and Likhoded \cite{BKL}.

In this paper, we study $D$-meson production under HERA and Tevatron
experimental conditions as well as the charm structure function $F_{2,c}$ of
the proton for the first time in the framework of the QMRK approach
\cite{QMRK1, QMRK2} complemented with the quark-Reggeization hypothesis.
This paper is organized as follows.
In Sec.~\ref{sec:two}, we present the basic formalism of our calculations and
briefly recall the QMRK approach in connection with the quark-Reggeization
hypothesis.
In Sec.~\ref{sec:three}, we consider the charm structure function $F_{2,c}$ of
the proton and compare our results with experimental data.
In Secs.~\ref{sec:four} and \ref{sec:five}, we describe $D$-meson production
via $c$-quark fragmentation at HERA and the Tevatron, respectively.
In Sec.~\ref{sec:six}, we summarize our conclusions.

\section{Basic formalism}
\label{sec:two}

In the phenomenology of the strong interactions at high energies, it is
necessary to describe the QCD evolution of the PDFs of the colliding particles
(hadrons or photons) starting from some scale $\mu_0$ which controls a
non-perturbative regime up to the typical scale $\mu$ of the hard-scattering
processes, which is typically of the order of the transverse mass
$M_T=\sqrt{M^2+p_T^2}$ of the produced particle (or hadron jet) with
(invariant) mass $M$ and transverse momentum $\vec{p}_T$.
In the region of very high energies, which corresponds to the so-called Regge
limit, the typical ratio $x=\mu/\sqrt{S}$ becomes very small, $x\ll1$.
This leads to large logarithmic contributions of the type
$[\alpha_s\ln(1/x)]^n$, where $\alpha_s$ is the strong-coupling constant,
which are conveniently resummed in the Balitsky-Fadin-Kuraev-Lipatov
\cite{BFKL} formalism by the evolution of unintegrated gluon and quark
distribution functions $\Phi_{g,q}^{p,\gamma}(x,q_T^2,\mu^2)$, where $x$ and
$\vec{q}_T$ are the longitudinal-momentum fraction and transverse momentum of
the Reggeized parton w.r.t.\ the parent particle, respectively.
Correspondingly, in the QMRK approach \cite{QMRK1,QMRK2}, the
initial-state $t$-channel gluons and quarks are considered as
Reggeons, or Reggeized gluons ($R$) and quarks ($Q$).
They carry finite transverse momenta $\vec{q}_T$ with respect to the hadron or
photon beam from which they stem and are off mass shell.

The advantages of the QMRK approach in comparison with the conventional
$k_T$-factorization scheme \cite{KTGribov} include:
firstly, it uses gauge-invariant amplitudes and is based on a factorization
hypothesis that is proven in the leading logarithmic approximation;
secondly, it carries over to non-leading orders in the strong-coupling
constant, as recently proven \cite{Fadinetal}.
The Reggeization of amplitudes provides the opportunity to efficiently take
into account large radiative corrections to processes in the Regge limit
beyond what is included in the collinear approximation, which is of great
practical importance.

Recently, the Feynman rules for the induced and some important effective
vertices of the effective theory based on the non-Abelian gauge-invariant
action \cite{KTLipatov} have been derived in Ref.~\cite{KTAntonov}.
However, these rules only refer to processes with Reggeized gluons in the
initial state.
As for $t$-channel quark-exchange processes, such rules are still unknown, so
that it is necessary to construct effective vertices involving Reggeized
quarks using QMRK approach prescriptions in each application from first
principles.
Of course, a certain set of Reggeon-Reggeon-Particle effective vertices are
known, for example for the transitions $RR\to g$ \cite{RRg},
$Q\overline{Q}\to g$ \cite{QQg}, and $RQ\to q$ \cite{BogdanFadin}.
The effective $\gamma^* Q\to q$ vertex, which describes the production of a
quark in the collision of a virtual photon with a Reggeized quark, has been
recently obtained in Ref.~\cite{SVA08}.

In our numerical calculations below, we adopt the prescription proposed by
Kimber, Martin, Ryskin, and Watt \cite{KMR} to obtain unintegrated gluon and
quark distribution functions for the proton from the conventional integrated
ones, as implemented in Watt's code \cite{code}.
To obtain the analogous unintegrated functions for the photon, we modify
Watt's code \cite{code}.
As input for this procedure, we use the Martin-Roberts-Stirlin-Thorne
\cite{MRST} proton and the Gl\"uck-Reya-Vogt \cite{GRV} photon PDFs.

\boldmath
\section{Charm structure function $F_{2,c}$ of the proton}
\label{sec:three}
\unboldmath

On the experimental side, the charm structure function $F_{2,c}$ of the proton
was measured by H1 \cite{Adloff02} and ZEUS \cite{Chekanov04} in deep
inelastic scattering (DIS) of electrons and positrons on protons at HERA.
In this section, we consider this quantity in the framework of the QMRK
approach endowed with the quark-Reggeization hypothesis.
We thus need the partonic cross section for the production of a $c$ quark in
the collision of a virtual photon and a Reggeized charm quark.
The relevant vertex was found in Ref.~\cite{SVA08} and reads:
\begin{equation}
C_{\gamma Q}^q=-ee_q\left[\frac{q_1^2}{q_1^2+q_2^2}\gamma^\mu
-\frac{2k^\mu}{q_1^2+q_2^2}\notq_2
+\frac{2x_2q_2^2P_2^\mu}{(q_1^2+q_2^2)^2}\notq_2\right],
\end{equation}
where the four-momenta of the virtual photon, the proton, the Reggeized charm
quark, and the outgoing charm quark are denoted as $q_1$, $P_2$,
$q_2=x_2P_2+q_{2T}$, and $k=q_1+q_2$, respectively.
We concentrate on photons with large virtuality $Q^2=-q_1^2\gg m_c^2$, so that
the massless approximation for describing DIS structure functions is
appropriate \cite{SVA08}.
We then obtain the following master formula for $F_{2,c}$:
\begin{equation}
F_{2,c}(x_B,Q^2)=2e_c^2\int_0^{Q^2}dt_2\,\Phi_c^p(x_2,t_2,\mu^2)
\frac{Q^2(Q^4+6Q^2t_2+2t_2^2)}{(Q^2+t_2)^3},
\end{equation}
where $e_c=2/3$ is the fractional electric charge of the $c$ quark and
$x_2=x_B(Q^2+t_2)/Q^2$, with $x_B$ being the Bjorken variable.
For definiteness, we choose the factorization scale to be $\mu^2=Q^2$.

\begin{figure}[ht]
\begin{center}
\begin{tabular}{ll}
\parbox{0.45\textwidth}{
\includegraphics[height=0.25\textheight,viewport=45 164 510 625,clip]{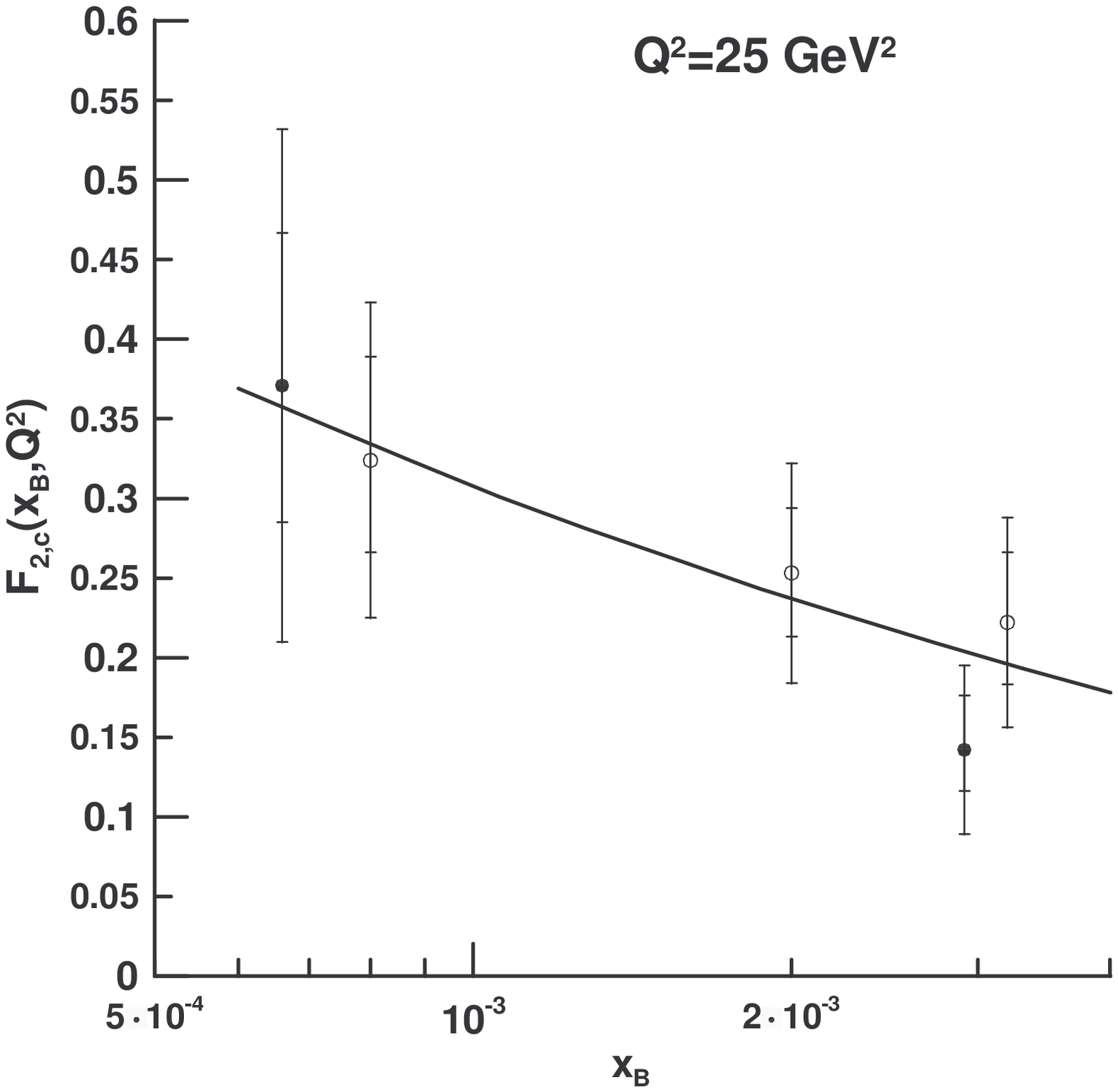}
}
&
\parbox{0.5\textwidth}{
\includegraphics[height=0.25\textheight,viewport=45 162 522 625,clip]{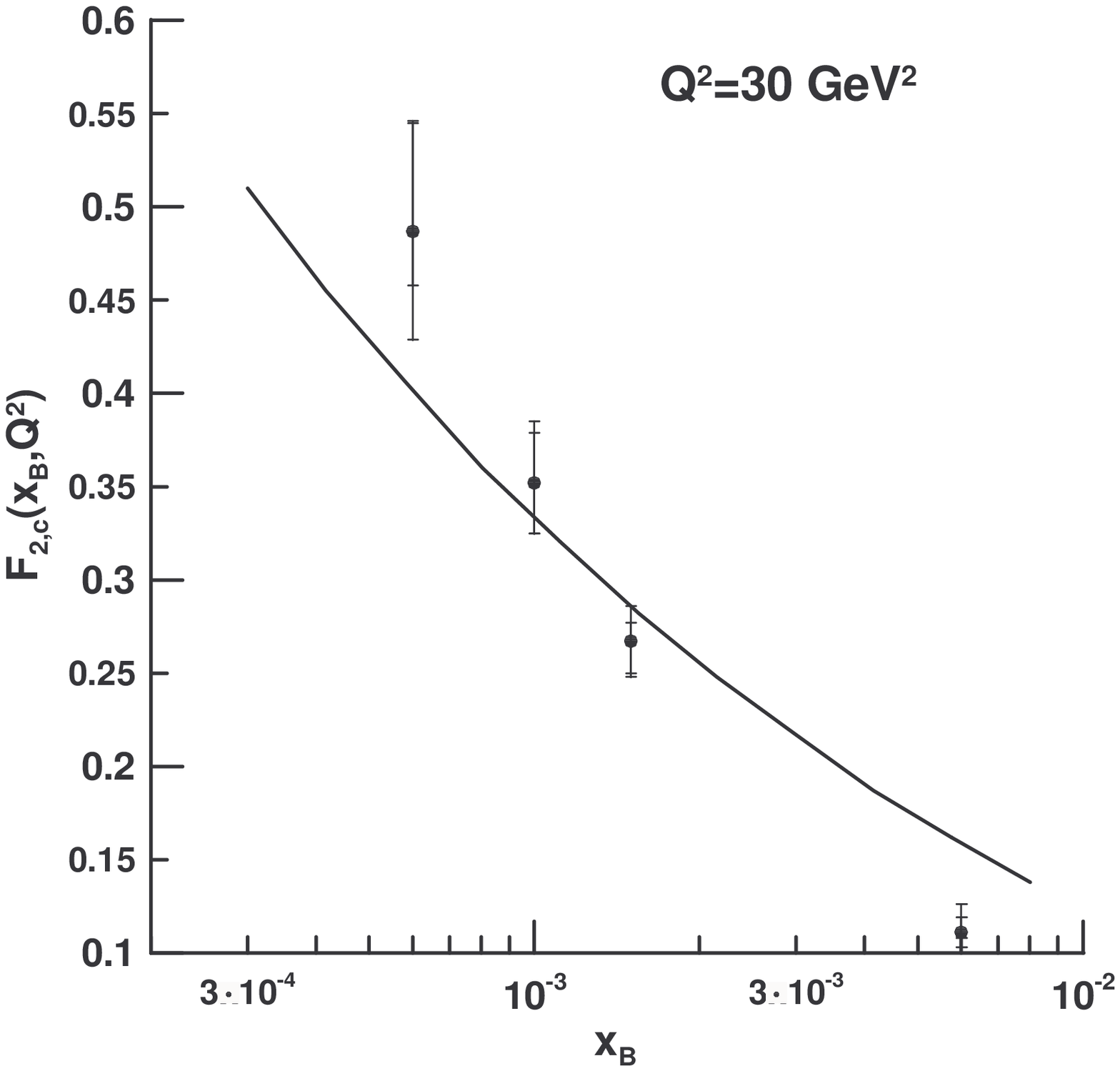}
}
\\
(a) & (b)\\
\parbox{0.5\textwidth}{
\includegraphics[height=0.25\textheight,viewport=51 162 522 623,clip]{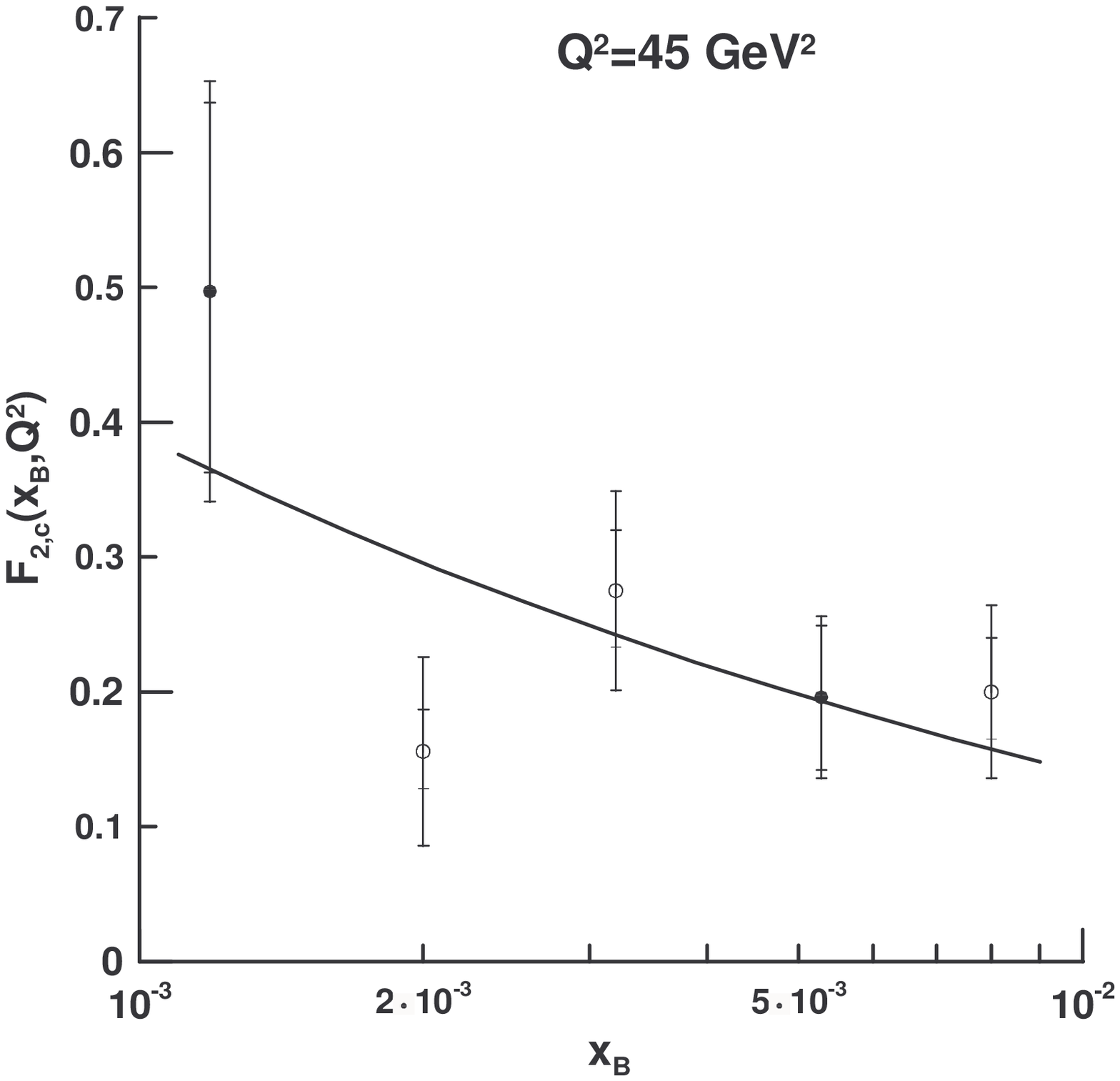}
}
&
\parbox{0.5\textwidth}{
\includegraphics[height=0.25\textheight,viewport=45 162 510 624,clip]{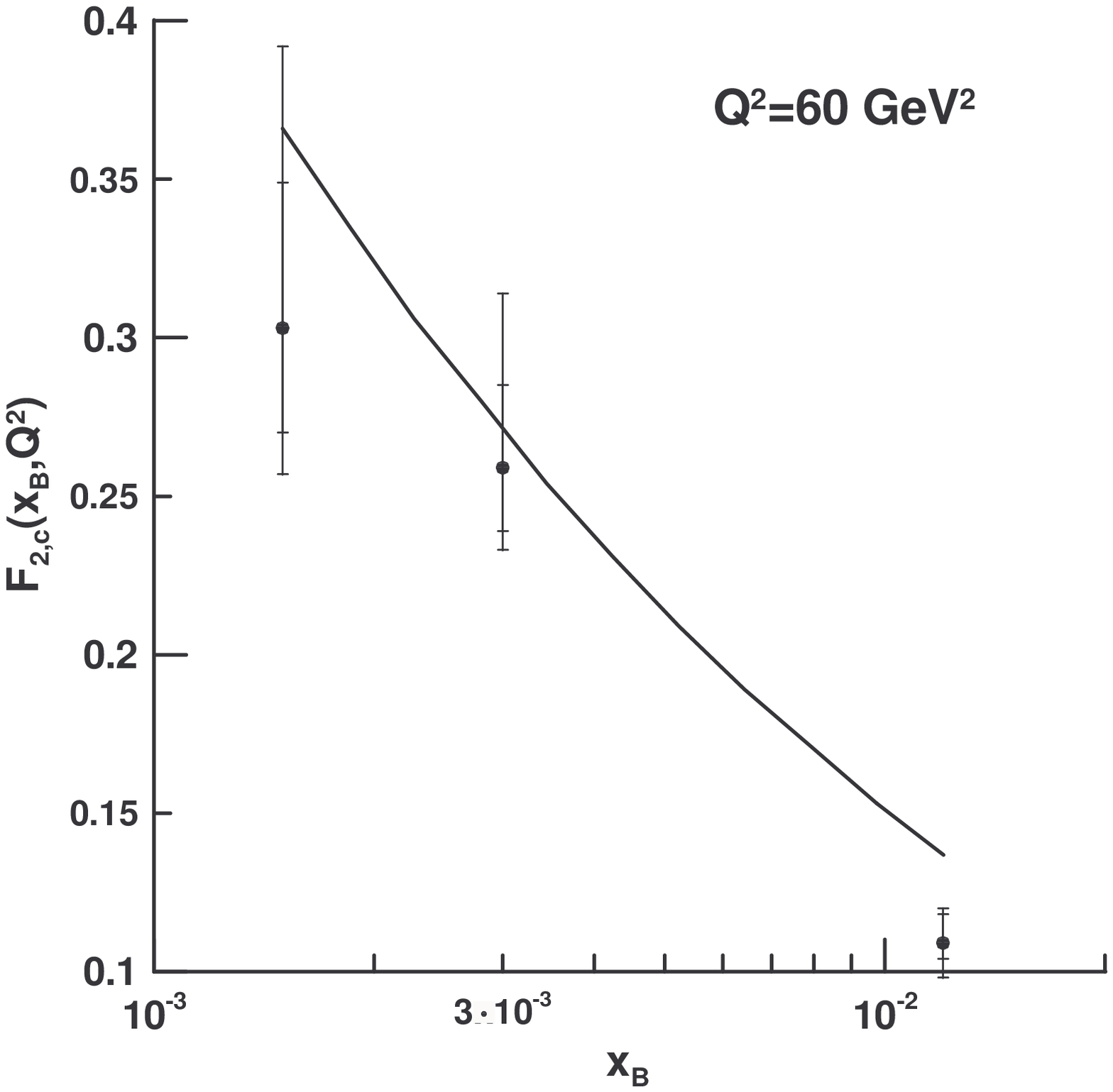}
}
\\
(c) & (d)\\
\parbox{0.5\textwidth}{
\includegraphics[height=0.25\textheight,viewport=45 162 510 625,clip]%
{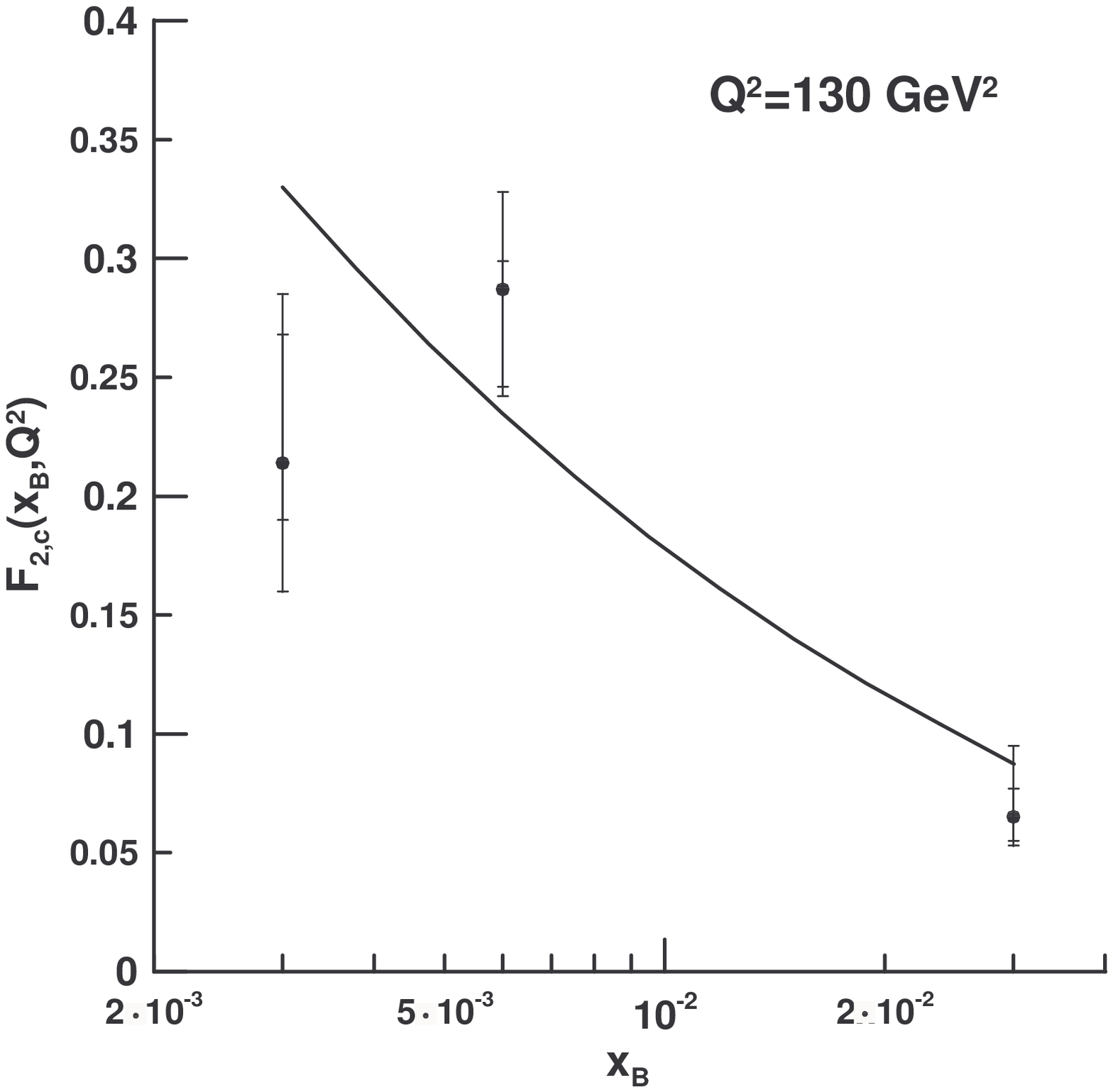}
}
&
\parbox{0.5\textwidth}{
\includegraphics[height=0.25\textheight,viewport=45 162 530 623,clip]%
{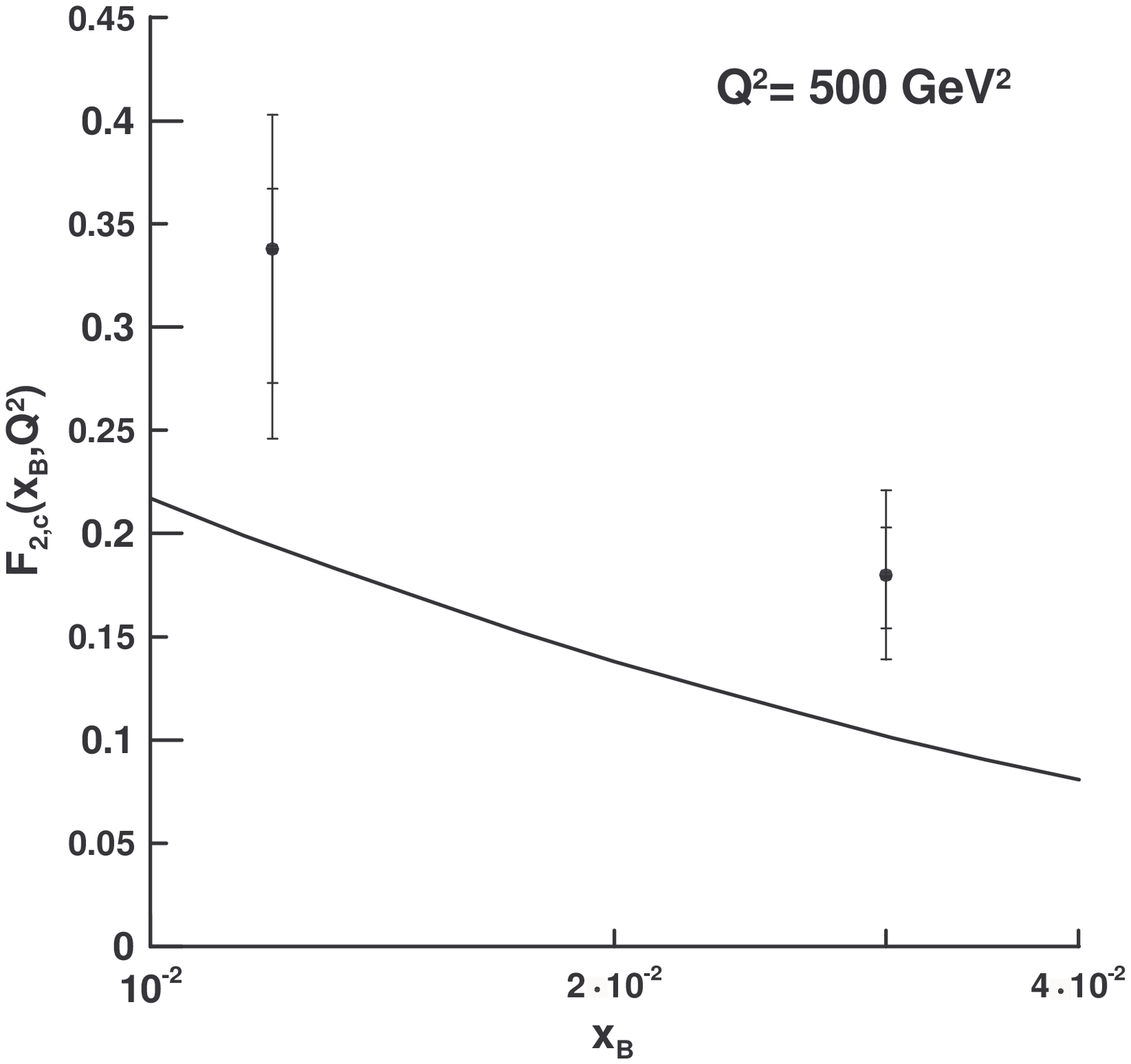}
}
\\
(e) & (f)
\end{tabular}
\end{center}
\caption{$F_{2,c}(x_B, Q^2)$ as a function of $x_B$ at (a) $Q^2=25$, (b) 30,
(c) 45, (d) 60, (e) 130, and (f) 500~GeV$^2$.
The H1 \cite{Adloff02} (open circles) and ZEUS \cite{Chekanov04} (filled
circles) are compared with LO predictions from the QMRK approach with the
quark-Reggeization hypothesis.}
\label{fig1}
\end{figure}

In Fig.~\ref{fig1}, we compare the $x_B$ distributions of $F_{2,c}$ for
various values of $Q^2$ with the H1 \cite{Adloff02} and ZEUS \cite{Chekanov04}
data.
We find good agreement for all values of $Q^2$, except for the highest one,
$Q^2=500$~GeV$^2$, where our prediction is about $50\%$ below the data.
This disagreement shows the importance of higher-order corrections at large
values of $Q^2$, which are beyond the scope of our present study.

\boldmath
\section{$D$-meson production at HERA}
\label{sec:four}
\unboldmath

On the experimental side, ZEUS measured the $p_T$ distributions of $D^{*\pm}$
\cite{HERADst} and $D_s^\pm$ \cite{HERADs} mesons with rapidity\footnote{%
Since we neglect finite quark and hadron mass effects, pseudorapidity and
rapidity coincide.}
$|y|\le1.5$
inclusively produced in photoproduction at HERA~I, with proton energy
$E_p=820$~GeV and lepton energy $E_e=27.5$~GeV in the laboratory frame, in the
ranges $2\le p_T\le 12$~GeV and $3\le p_T\le 12$~GeV, respectively.
In this section, we compare this data with our QMRK predictions.
At leading order (LO), we need to consider only three $2\to 1$ partonic
subprocesses, namely $C_p\gamma\to c$ for direct photoproduction and
$C_p R_\gamma\to c$ and $R_p C_\gamma\to c$ for resolved photoproduction, where
the subscript indicates the mother particle.

Exploiting the factorization theorem, the $p_T$ distribution of direct
photoproduction takes the form
\begin{eqnarray}
p_T^3\frac{d\sigma}{dp_T}&=&2\pi\int dy\int dz\,
x_\gamma f_{\gamma/e}(x_\gamma)
z^2 D_{c\to D}(z,\mu^2)
\nonumber\\
&&{}\times\Phi_{c}^p(x_1,t_1,\mu^2)\overline{|M(C_p\gamma\to c)|^2},
\label{eq:dir}
\end{eqnarray}
where
\begin{equation}
x_1=\frac{p_T e^y}{2zE_p},\qquad
x_\gamma=\frac{p_T e^{-y}}{2zE_e},\qquad
t_1=k_T^2,\qquad
\vec{k}_T=\frac{\vec{p}_T}{z},
\end{equation}
with $\vec{k}_T$ being the transverse momentum of the produced $c$ quark.
We evaluate the quasi-real-photon flux $f_{\gamma/e}$ in
Weizs\"acker-Williams approximation using
\begin{equation}
f(x_\gamma)=\frac\alpha{2\pi}\left[\frac{1+(1-x_\gamma)^2}{x_\gamma}
\ln\frac{Q_{\rm max}^2}{Q_{\rm min}^2}+
2m_e^2x_\gamma\left(\frac{1}{Q^2_{\rm min}}-\frac{1}{Q^2_{\rm max}}\right)
\right],
\end{equation}
where $\alpha$ is Sommerfeld's fine-structure constant, $m_e$ is the electron
mass, $Q^2_{\rm min}=m_e^2x_\gamma^2/(1-x_\gamma)$, and $Q^2_{\rm max}$ is
determined by the experimental setup, with $Q^2_{\rm max}=1$~GeV$^2$ in our
case \cite{HERADst,HERADs}.
As for the $c\to D$ fragmentation function (FF) $D_{c\to D}$, we adopt the
non-perturbative $D^{*\pm}$ and $D_s^\pm$ sets determined in the ZM-VFNS with
initial evolution scale $\mu_0=m_c$ \cite{FF} from fits to OPAL data from CERN
LEP1.
We choose the renormalization and initial- and final-state factorization
scales to be $\mu=\sqrt{m_D^2+p_T^2}$, where $m_D$ is the $D$-meson mass.
Using the Reggeized-quark--photon effective vertex from Ref.~\cite{SVA08},
the square of the hard-scattering amplitude is found to be
\begin{equation}
\overline{|M(C\gamma\to c)|^2}=8\pi\alpha e_c^2k_T^2.
\end{equation}
It is understood that also the contribution from the charge-conjugate partonic
subprocess is to be included in Eq.~(\ref{eq:dir}).

In the case of resolved photoproduction via the partonic subprocess
$C_p R_\gamma\to c$, the factorization formula reads:
\begin{eqnarray}
p_T^3\frac{d\sigma}{dp_T}&=&\int dy\int dz\int dx_\gamma\int dt_2\int d\phi_2
\, f_{\gamma/e}(x_\gamma)z^2D_{c\to D}(z,\mu^2)\nonumber\\
&&{}\times\Phi_{c}^p(x_1,t_1,\mu^2)\Phi_g^\gamma(x_2,t_2,\mu^2)
\overline{|M(C_p R_\gamma \to c)|^2},
\label{eq:res}
\end{eqnarray}
where
\begin{equation}
x_1=\frac{p_T e^y}{2zE_p},\qquad
x_2=\frac{p_T e^{-y}}{2x_\gamma zE_e},\qquad
t_1=t_2-2k_T\sqrt{t_2}\cos\phi_2+k_T^2,\qquad
t_2=q_{2T}^2,\qquad
\vec{k}_T=\frac{\vec{p}_T}{z},
\end{equation}
with $\phi_2$ being the angle enclosed between $\vec{p}_T$ and $\vec{q}_{2T}$.
Using the Reggeized-quark--Reggeized-gluon effective vertex from
Ref.~\cite{BogdanFadin}, we have
\begin{equation}
\overline{|M(CR\to c)|^2}=\frac{2}{3}\pi\alpha_s(\mu^2)k_T^2.
\label{eq:ms}
\end{equation}
Again, the charge-conjugate partonic subprocess is to be included in
Eq.~(\ref{eq:res}).
Resolved photoproduction via the partonic subprocess $R_p C_\gamma\to c$ is
treated very similarly.

\begin{figure}[ht]
\begin{center}
\begin{tabular}{ll}
\parbox{0.5\textwidth}{
\includegraphics[width=0.5\textwidth,viewport=58 182 505 608,clip]%
{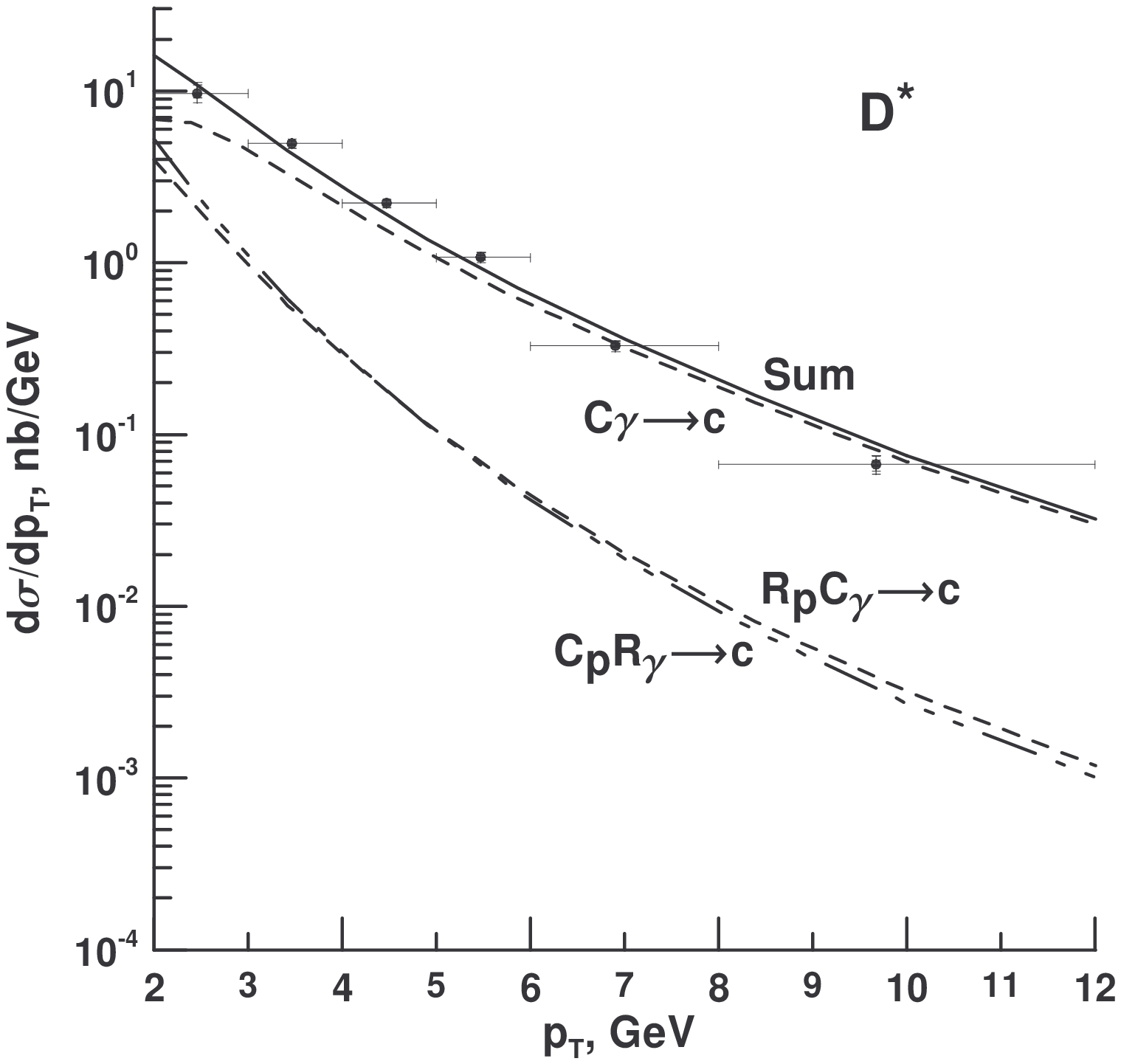}
}
&
\parbox{0.5\textwidth}{
\includegraphics[height=0.5\textwidth,viewport=58 182 505 608,clip]%
{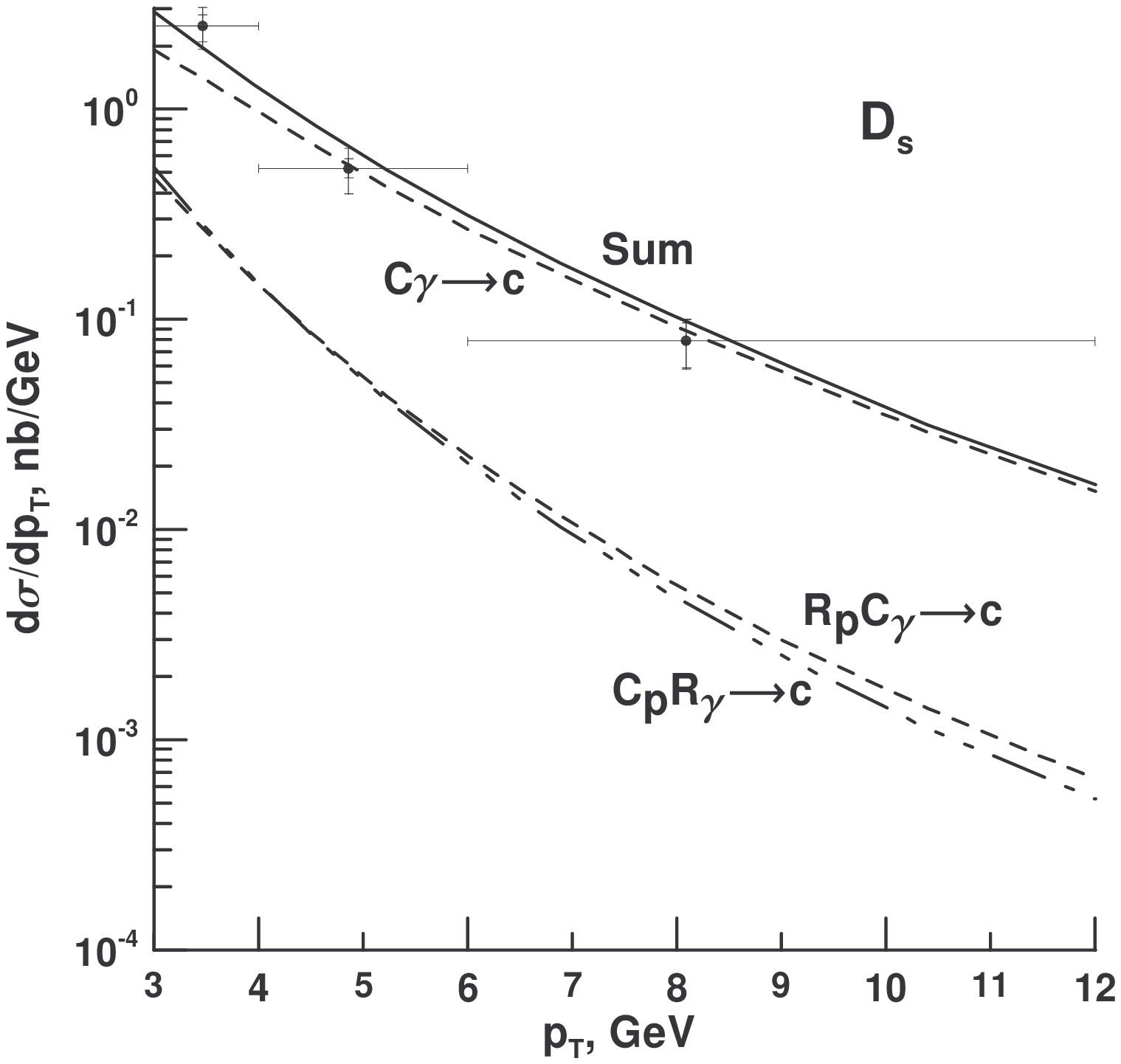}
}
\\
(a) & (b)
\end{tabular}
\end{center}
\caption{$p_T$ distributions of inclusive (a) $D^{*\pm}$ and (b) $D_s^\pm$
photoproduction for $\sqrt S=300$~GeV and $|y|\le1.5$.
The ZEUS data from (a) Ref.~\cite{HERADst} and (b) Ref.~\cite{HERADs} are
compared with LO predictions from the QMRK approach with the
quark-Reggeization hypothesis.}
\label{figDHERA}
\end{figure}

In Figs.~\ref{figDHERA}(a) and (b), our results for $D^{*\pm}$ and $D_s^\pm$
mesons, respectively, are broken down to the $C_p\gamma\to c$,
$C_p R_\gamma\to c$, and $R_p C_\gamma\to c$ contributions and are compared
with the ZEUS data \cite{HERADst,HERADs}.
We find that the theoretical predictions are dominated by direct
photoproduction and agree rather well with the experimental data over the
whole $p_T$ range considered.

\boldmath
\section{$D$-meson production at the Tevatron}
\label{sec:five}
\unboldmath

CDF \cite{DTevn} measured the $p_T$ distributions of $D^0$, $D^\pm$,
$D^{*\pm}$, and $D_s^\pm$ mesons with rapidity $|y|\le1$ inclusively produced
in hadroproduction in run~II at the Tevatron, with $\sqrt S=1.96$~TeV.
To LO in the QMRK approach, the factorization formula for the
$C_pR_{\overline{p}}\to c$ channel reads:
\begin{eqnarray}
p_T^3\frac{d\sigma}{dp_T}&=&\int dy\int dz \int dt_1 \int d\phi_1\,
z^2 D_{c\to D}(z,\mu^2)\nonumber\\
&&{}\times\Phi_{c}^p(x_1,t_1,\mu^2)
\Phi_g^{\overline{p}}(x_2,t_2,\mu^2)
\overline{|M(C_p R_{\overline{p}}\to c)|^2},
\label{eq:had}
\end{eqnarray}
where $\overline{|M(C_p R_{\overline{p}} \to c)|^2}$ is given by
Eq.~(\ref{eq:ms}),
\begin{equation}
x_1=\frac{p_T e^y}{z\sqrt S},\qquad
x_2=\frac{p_T e^{-y}}{z\sqrt S},\qquad
t_2=t_1-2\frac{p_T}z\sqrt{t_1}\cos\phi_2+\frac{p_T^2}{z^2}.
\end{equation}
The result for the $R_pC_{\overline{p}}\to c$ channel is similar and has to
be included in Eq.~(\ref{eq:had}) together with those from the
charge-conjugate partonic subprocesses.

\begin{figure}[ht]
\begin{center}
\begin{tabular}{ll}
\parbox{0.5\textwidth}{
\includegraphics[width=0.5\textwidth,viewport=60 182 506 615,clip]{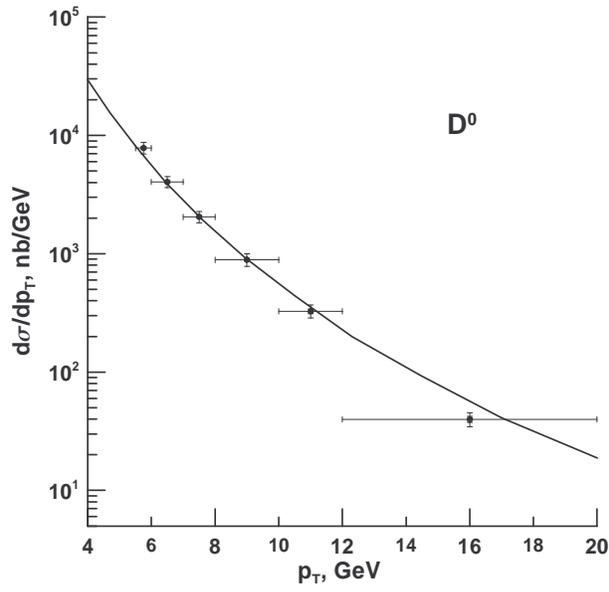}
}
&
\parbox{0.5\textwidth}{
\includegraphics[height=0.5\textwidth,viewport=60 182 506 615,clip]{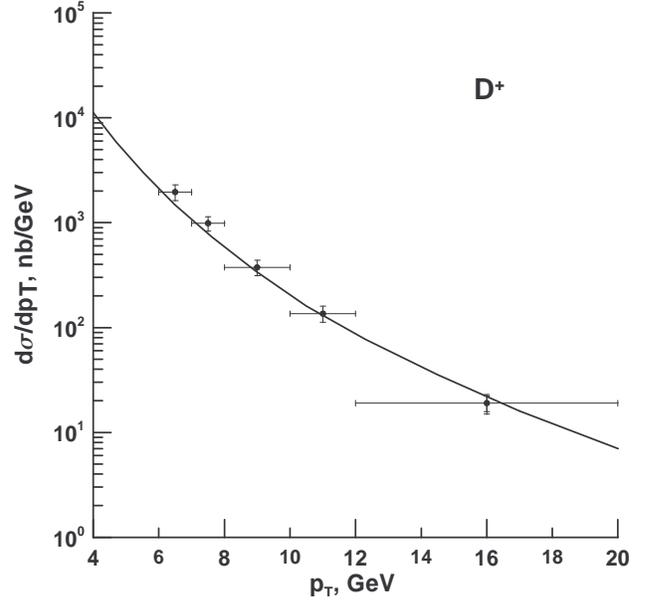}
}
\\
(a) & (b)\\
\parbox{0.5\textwidth}{
\includegraphics[width=0.5\textwidth,viewport=48 169 519 619,clip]{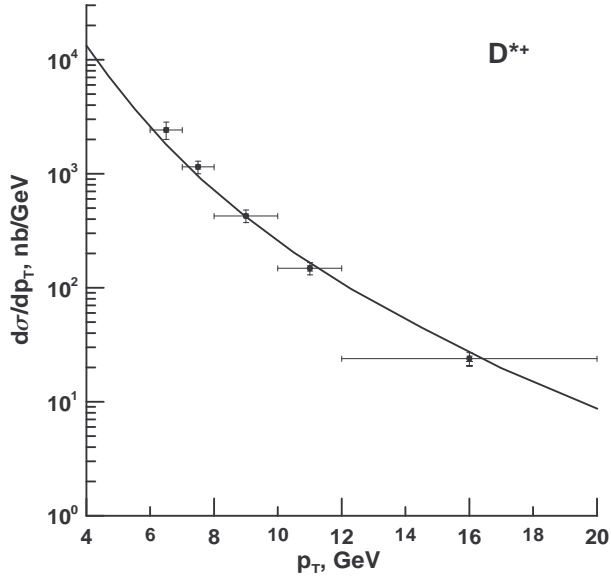}
}
&
\parbox{0.5\textwidth}{
\includegraphics[width=0.5\textwidth,viewport=48 169 519 626,clip]{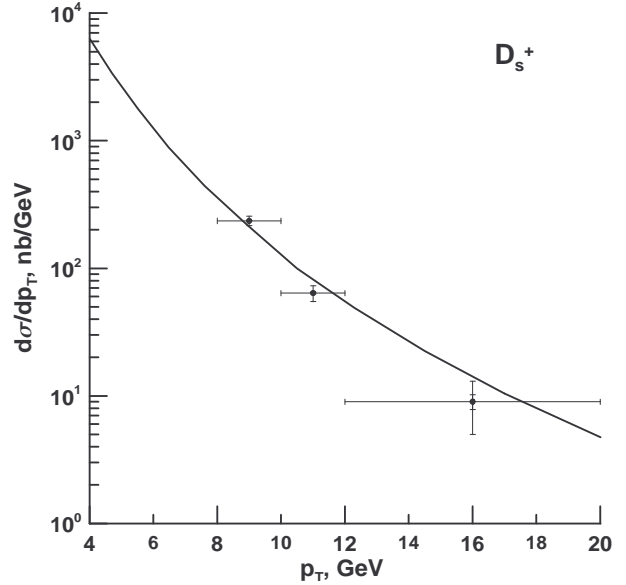}
}
\\
(c) & (d)
\end{tabular}
\end{center}
\caption{$p_T$ distributions of inclusive (a) $D^0$, (b) $D^\pm$, (c)
$D^{*\pm}$, and (d) $D_s^\pm$ hadroproduction for $\sqrt S=1.96$~TeV and
$|y|\le1$.
The CDF data from Ref.~\cite{DTevn} are compared with LO predictions from the
QMRK approach with the quark-Reggeization hypothesis.}
\label{figDTevn}
\end{figure}

In Figs.~\ref{figDTevn}(a)--(d), our results for $D^0$, $D^\pm$, $D^{*\pm}$,
and $D_s^\pm$ mesons, respectively, are compared with the CDF data
\cite{DTevn}.
We find that the theoretical predictions generally agree rather well with the
experimental data, except perhaps for the slope.
In fact, the predictions exhibit a slight tendency to undershoot the data at
small values of $p_T$ and to overshoot them at large values of $p_T$.
However, we have to bear in mind that these are just LO predictions, so that
there is room for improvement by including higher orders.

In the framework of the collinear parton model, comparisons with the
experimental data of Ref.~\cite{DTevn} were performed beyond LO, namely in the
fixed-order-next-to-leading-logarithm (FONLL) scheme \cite{Nason} and at NLO
in the GM-VFNS \cite{KKSS,Kniehl:2008eu}.
The FONLL predictions systematically undershoot the CDF data \cite{DTevn}.
The GM-VFNS predictions of Ref.~\cite{KKSS}, which are evaluated with FFs
determined in the ZM-VFNS \cite{FF}, describe these data within their
errors, but are still somewhat on the low side.
The degree of agreement is further improved \cite{Kniehl:2008eu} by evaluating
the GM-VFNS predictions of Ref.~\cite{KKSS} using FFs extracted
\cite{Kneesch:2007ey} from a global fit to $B$- and $Z$-factory data of
$e^+e^-$ annihilation in the very same scheme.

\section{Conclusions}
\label{sec:six}

In this paper, we explored the usefulness of the quark-Reggeization hypothesis
in the framework of the QMRK approach by studying several observables of
inclusive charm production at LO, namely the charm structure function
$F_{2,c}$ of the proton measured at HERA \cite{Adloff02,Chekanov04} as well as
the one-particle-inclusive cross sections of $D^{*\pm}$ and $D_s^\pm$
photoproduction in $ep$ collisions at HERA \cite{HERADst,HERADs} and of $D^0$,
$D^\pm$, $D^{*\pm}$, and $D_s^\pm$ hadroproduction in $p\overline{p}$
collisions at the Tevatron \cite{DTevn}.
In all three cases, we found satisfactory agreement between our default
predictions and the experimental data, which is quite encouraging in view of
the simplicity of our LO expressions for the partonic cross sections.
By contrast, in the collinear parton model of QCD, the inclusion of NLO
corrections is necessary to achieve such a degree of agreement.
We thus recover the notion that the QMRK approach is a powerful tool for the
theoretical description of QCD processes in the high-energy limit and
automatically accommodates an important class of corrections that lie beyond
the reach of the collinear parton model at LO \cite{PRD2003}.

\section*{Acknowledgments}

We thank L.~N.~Lipatov and O.~V.~Teryaev for useful discussions.
A.~V.~S. is grateful to the International Center of Fundamental Physics in
Moscow and to the Dynastiya Foundation for financial support.
This work was supported in part by the German Federal Ministry for Education
and Research BMBF through Grant No.\ 05~HT6GUA and by the German Research
Foundation DFG through Grant No.\ KN~365/7--1

\end{document}